\documentclass[12pt]{article}

\usepackage{amsthm,amsmath,amsfonts,amssymb}
\usepackage{graphicx}

\usepackage{natbib}
\usepackage{times}
\usepackage{bm}
\usepackage{bbm}
\usepackage{tikz}
\usepackage{color}
\usepackage{multirow}
\usepackage{float}
\usepackage{enumitem}
\usepackage{booktabs,caption,subcaption}
\usepackage{titlesec}

\usepackage[margin=1in]{geometry}
\allowdisplaybreaks
\usepackage{setspace}
\newtheorem{theorem}{Theorem}[section]

\theoremstyle{remark}

\newcommand{\bA}{\boldsymbol{A}}
\newcommand{\ba}{\boldsymbol{a}}
\newcommand{\bB}{\boldsymbol{B}}

\newcommand{\bC}{\boldsymbol{C}}

\newcommand{\bD}{\boldsymbol{D}}

\newcommand{\bJ}{\boldsymbol{J}}
\newcommand{\bj}{\boldsymbol{j}}

\newcommand{\bS}{\boldsymbol{S}}
\newcommand{\bs}{\boldsymbol{s}}

\newcommand{\bV}{\boldsymbol{V}}
\newcommand{\bv}{\boldsymbol{v}}

\newcommand{\bW}{\boldsymbol{W}}

\newcommand{\bX}{\boldsymbol{X}}
\newcommand{\bx}{\boldsymbol{x}}
\newcommand{\bY}{\boldsymbol{Y}}
\newcommand{\by}{\boldsymbol{y}}

\newcommand{\bmu}{\boldsymbol{\mu}}

\newcommand{\bPsi}{\boldsymbol{\Psi}}
\newcommand{\bpsi}{\boldsymbol{\psi}}

\newcommand{\bSigma}{\boldsymbol{\Sigma}}

\newcommand{\btheta}{\boldsymbol{\theta}}

\begin{document}

\begin{singlespace}

\title{\bf Joint integrative analysis of multiple data sources with correlated vector outcomes}

\author{Emily C. Hector \\Department of Statistics\\North Carolina State University\\
and\\
Peter X.-K. Song\\Department of Biostatistics\\University of Michigan}

\date{}

\maketitle

\begin{abstract}
We propose a distributed quadratic inference function framework to jointly estimate regression parameters from multiple potentially heterogeneous data sources with correlated vector outcomes. The primary goal of this joint integrative analysis is to estimate covariate effects on all outcomes through a marginal regression model in a statistically and computationally efficient way. We develop a data integration procedure for statistical estimation and inference of regression parameters that is implemented in a fully distributed and parallelized computational scheme. To overcome computational and modeling challenges arising from the high-dimensional likelihood of the correlated vector outcomes, we propose to analyze each data source using \cite{Qu-Lindsay-Li}'s quadratic inference functions, and then to jointly reestimate parameters from each data source by accounting for correlation between data sources using a combined meta-estimator in a similar spirit to \cite{Hansen}'s generalised method of moments. We show both theoretically and numerically that the proposed method yields efficiency improvements and is computationally fast. We illustrate the proposed methodology with the joint integrative analysis of the association between smoking and metabolites in a large multi-cohort study and provide an R package for ease of implementation. 
\end{abstract}

\noindent%

\end{singlespace}

\vfill

\newpage

\section{Introduction}

\begin{figure}[H]
\caption{Pearson correlation of peptide metabolites observed in four cohorts, with sub-pathways denoted by tick marks. \label{figure-data-correlation-studies}}
\resizebox{\textwidth}{!}{
\begin{tikzpicture}[every node/.style={anchor=north west,inner sep=0pt}, x=1mm, y=1mm]   
\node (image) at (0,0) {\includegraphics[scale=1.5]{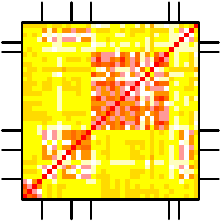}};
\node (image) at (35,0) {\includegraphics[scale=1.5]{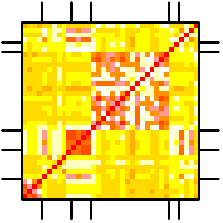}};
\node (image) at (70,0) {\includegraphics[scale=1.5]{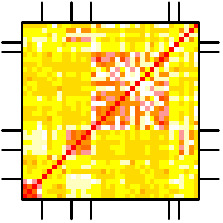}};
\node (image) at (105,0) {\includegraphics[scale=1.5]{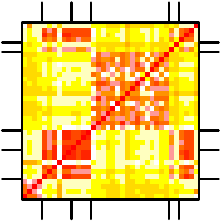}};
\node (image) at (0,-35) {\includegraphics[width=13.9cm]{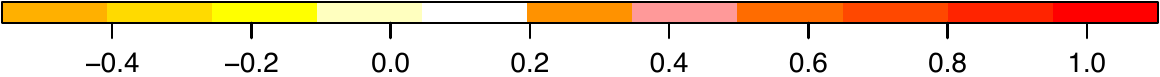}};
\end{tikzpicture}
}
\end{figure}

Data integration methods have drawn increasing attention with the availability of massive data from multiple sources, with proposed methods spanning the gamut from the frequentist confidence distribution approach \citep{Xie-Singh-Strawderman, Xie-Singh} to Bayesian hierarchical models \citep{Smith-Spiegelhalter-Thomas}, as well as several generalisations of \cite{Glass}'s meta-analysis \citep{Ioannidis, DerSimonian-Laird, Kundu-Tang-Chatterjee}.  This paper is substantially motivated by the analysis of the effect of smoking on metabolites that are upstream determinants of cardiovascular health. We consider the analysis of four independent cohorts that quantify metabolites across multiple dependent metabolic sub-pathways in the Metabolic Syndrome in Men (METSIM) study \citep{Laakso-etal}. We refer to each cohort and sub-pathway as a data source. Of interest is a pathway-specific joint integrative regression analysis of all independent cohorts and dependent sub-pathways in the pathway. As visualized in Figure \ref{figure-data-correlation-studies} for the peptide pathway, metabolites in a pathway have heterogeneous correlation structures within and between data sources; see Supplementary Material \citep{Hector-Song-DIQIF} for all pathways. Developing a joint integrative regression framework that incorporates this complex correlation structure is key in obtaining improvements in estimation efficiency that yield new scientific findings. We propose a distributed quadratic inference function framework for this joint integrative regression analysis that addresses five major aspects of data integration: correlation of data sources, heterogeneity of data sources, statistical efficiency, privacy concerns and computational speed.\\
Recent work has primarily focused on synthesizing evidence from independent data sources, as in \cite{Claggett-Xie-Tian} and \cite{Yang-etal}. In practice, however, studies may collect correlated outcomes from different structural modalities, such as high-dimensional longitudinal phenotypes, pathway-networked omics biomarkers, or brain imaging measurements, which collectively form one high-dimensional correlated response vector for each participant. Of interest is conducting inference integrated not only over the independent data sources but also over the structurally correlated outcomes. High-order moments of complex high-dimensional correlated data may be difficult to model or handle computationally, which has led many to use working independence assumptions at the cost of statistical efficiency, resulting in potentially misleading statistical inference; see for example the composite likelihood approach in \cite{Caragea-Smith} and \cite{Varin}. \cite{Hector-Song-JMLR} proposes a method to account for correlation between data sources without specifying a full parametric model, but their method is burdened by the estimation of a high-dimensional parameter related to the second-order moments, whose dimension can rapidly increase and exceed the sample size as the number of data sources increases. To relieve this burden, we propose a fast and efficient approach that avoids estimation of parameters in  second-order moments with no loss of statistical efficiency.\\
Traditional data integration methods, such as meta-analysis and the confidence distribution approach, frequently assume parameter or even likelihood homogeneity across data sources, which often does not hold in practice. Data source heterogeneity can stem from differences in populations, study design, or associations, and can result in first- and higher-order moment heterogeneity. On the other hand, seemingly unrelated regression \citep{Zellner} can be inefficient when some parameters are homogeneous. One approach to dealing with first-order moment heterogeneity is to include data source-specific random effects, which can be inefficient and may induce misspecified correlation. Another approach is to allow cohort-specific fixed-effects, as in \cite{Lin-Zeng, Liu-Liu-Xie, Hector-Song-JMLR}. In the current literature there is a lack of computationally fast and statistically efficient methods to handle high-dimensional second- and higher-order moment parameters, which are regarded as nuisance parameters in a correlated data integrative setting. With only one data source, the quadratic inference function \citep{Qu-Lindsay-Li} is widely used to estimate regression parameters in first-order moments while avoiding estimation of second- and higher-order moments. Thus, the quadratic inference function minimizes the excessive burden of handling nuisance parameters. Our proposed distributed quadratic inference functions estimate regression parameters in mean models for each data source, thereby avoiding estimation of nuisance parameters in higher-order moments, and linearly updates the regression parameters according to different heterogeneity patterns across data sources. Not only does our approach combine the strengths from both meta-analysis and seemingly unrelated regression, but it is more flexible than these two methods. \\
For privacy reasons we may not have access to individual level data when integrating correlated data sources, in which case it becomes imperative to develop methods that can be implemented in a computationally distributed fashion. Even with access to individual level data, distributed algorithms are often preferred for their ability to significantly reduce the computational burden of traditional inference methods \citep{Jordan, Fan-Han-Liu}. There is a need for distributed methods able to handle parameter heterogeneity for computationally and statistically efficient inference with multiple correlated data sources.\\
Our proposed distributed quadratic inference function approach estimates mean parameters for cohort- and outcome-specific models in the integrative analysis of correlated outcome vectors while avoiding estimation of second-order moments. It yields statistically efficient estimation within a broad class of models. Cohort- and outcome-specific models are then selectively combined via a meta-estimator similar in spirit to \cite{Hansen}'s generalised method of moments according to some characterization of data heterogeneity. This new method has two major advantages over existing methods: the integrated estimator does not require access to individual-level data, and it can be computed non-iteratively to minimise computational costs.\\
The paper is organized as follows. In Section \ref{sec:METSIM} we briefly describe the motivating METSIM study and regression analysis problem. In Section \ref{sec:method} we describe the proposed joint integrative regression method. We study the numerical performance of our proposed method through simulations in Section \ref{sec:simulations}. The integrative analysis of metabolite sub-pathways in METSIM is given in Section \ref{sec:data}. We conclude with a discussion in Section \ref{sec:discussion}. Theoretical justifications and additional numerical results are deferred to the Appendices and Supplemental Material \citep{Hector-Song-DIQIF} respectively.

\section{Metabolic Syndrome in Men study}
\label{sec:METSIM}

The Metabolic Syndrome in Men study is a population-based study of 10197 Finnish men with the aim of investigating nongenetic and genetic factors associated with the risk of Type 2 diabetes, cardiovascular disease, and cardiovascular risk factors \citep{Laakso-etal}. The Centers for Disease Control and Prevention list smoking as a major cause of cardiovascular disease and the 2014 Surgeon General's Report on smoking and health reported that smoking was responsible for one of every four deaths from cardiovascular disease. Investigating the association between smoking and metabolites can provide insight into the etiology of metabolic diseases such as cardiovascular disease. \\
Using the Metabolon platform, the Metabolic Syndrome in Men study measured metabolites in eight pathways in four separate samples of men. Of interest is a regression analysis of the effect of smoking on metabolites in each of the eight pathways. Due to the complex correlation structure within and between sub-pathways in a pathway, we focus on a local model specification within each cohort and sub-pathway, termed a data source, and aim to integrate models across data sources for a joint integrative regression approach. For each pathway, the effect of smoking is known \textit{a priori} to be partially homogeneous across sub-pathways: the regression models for the sub-pathways can be grouped according to a known partition scheme $\mathcal{P}$ that describes the homogeneity/heterogeneity pattern of regression coefficients from different sub-pathways. For example, as illustrated in Figure \ref{peptide-partition}, the effect of smoking on peptide metabolites is homogeneous across the gamma-glutamyl amino acid, modified peptide, and polypeptide sub-pathways, and heterogeneous across the acetylated peptide, dipeptide and fibrinogen cleavage peptide sub-pathways. The dimension of the regression parameter of interest therefore is a function both of the number of covariates and the number of heterogeneous partitions. \\
In Section \ref{sec:method} we describe the general framework for estimating pathway-specific regression parameters that are partially homogeneous across the independent cohorts and dependent sub-pathways. For this general framework, we focus on one pathway and refer to each cohort and sub-pathway as a data source. The application of the proposed method to the METSIM study is revisited in Section \ref{sec:data}.

\section{Distributed and integrated quadratic inference functions}
\label{sec:method}

\subsection{Model formulation}
\label{subsec:formulation}
\begin{figure}[h!]
\caption{Pearson correlation of peptide metabolites observed in four cohorts and six sub-pathways: acetylated peptides (SP1), dipeptide (SP2), fibrinogen cleavage peptide (SP3), gamma-glutamyl amino acid (SP4), modified peptides (SP5), and polypeptide (SP6). Partition $\mathcal{P}=\{\mathcal{P}_g \}_{g=1}^4$ is contoured in blue dashed rectangles. \label{peptide-partition}}
\centering
\begin{tikzpicture}[every node/.style={anchor=north west,inner sep=0pt}, x=1mm, y=1mm]   
\node (image) at (0,0) {\includegraphics[scale=0.7]{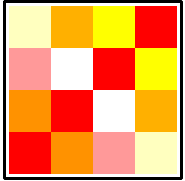}};
\node (image) at (20,0) {\includegraphics[scale=0.7]{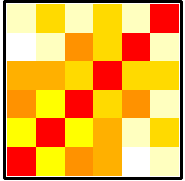}};
\node (image) at (40,0) {\includegraphics[scale=0.7]{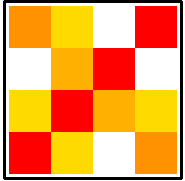}};
\node (image) at (60,0) {\includegraphics[scale=0.7]{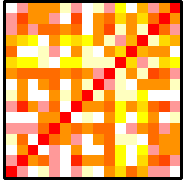}};
\node (image) at (80,0) {\includegraphics[scale=0.7]{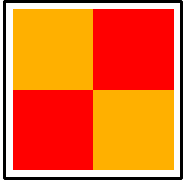}};
\node (image) at (100,0) {\includegraphics[scale=0.7]{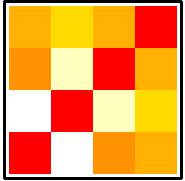}};

\node (image) at (0,-15) {\includegraphics[scale=0.7]{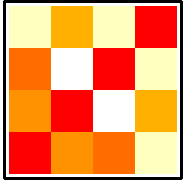}};
\node (image) at (20,-15) {\includegraphics[scale=0.7]{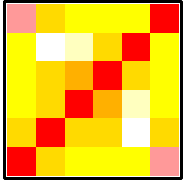}};
\node (image) at (40,-15) {\includegraphics[scale=0.7]{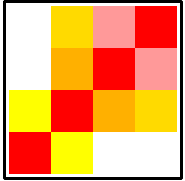}};
\node (image) at (60,-15) {\includegraphics[scale=0.7]{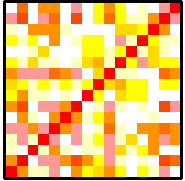}};
\node (image) at (80,-15) {\includegraphics[scale=0.7]{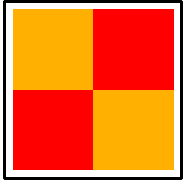}};
\node (image) at (100,-15) {\includegraphics[scale=0.7]{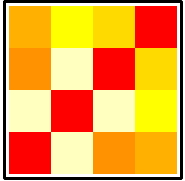}};

\node (image) at (0,-30) {\includegraphics[scale=0.7]{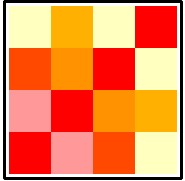}};
\node (image) at (20,-30) {\includegraphics[scale=0.7]{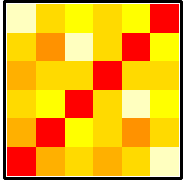}};
\node (image) at (40,-30) {\includegraphics[scale=0.7]{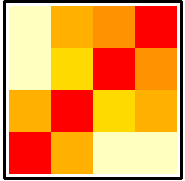}};
\node (image) at (60,-30) {\includegraphics[scale=0.7]{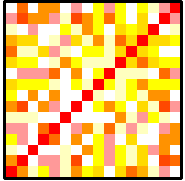}};
\node (image) at (80,-30) {\includegraphics[scale=0.7]{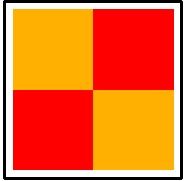}};
\node (image) at (100,-30) {\includegraphics[scale=0.7]{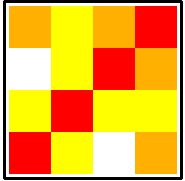}};

\node (image) at (0,-45) {\includegraphics[scale=0.7]{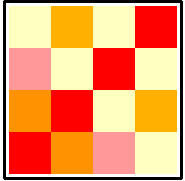}};
\node (image) at (20,-45) {\includegraphics[scale=0.7]{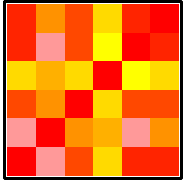}};
\node (image) at (40,-45) {\includegraphics[scale=0.7]{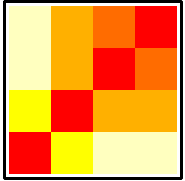}};
\node (image) at (60,-45) {\includegraphics[scale=0.7]{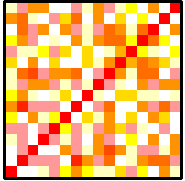}};
\node (image) at (80,-45) {\includegraphics[scale=0.7]{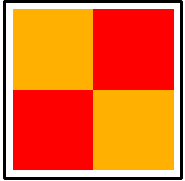}};
\node (image) at (100,-45) {\includegraphics[scale=0.7]{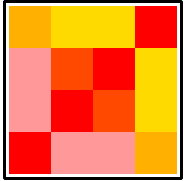}};

\draw[blue,dashed,line width=0.75mm] (-1,1) rectangle (14,-59);
\draw[blue,dashed,line width=0.75mm] (19,1) rectangle (34,-59);
\draw[blue,dashed,line width=0.75mm] (39,1) rectangle (54,-59);
\draw[blue,dashed,line width=0.75mm] (59,1) rectangle (114,-59);
\node[font={\normalsize}] at (-20,-1) {Cohort 1};
\node[font={\normalsize}] at (-20,-16) {Cohort 2};
\node[font={\normalsize}] at (-20,-31) {Cohort 3};
\node[font={\normalsize}] at (-20,-45) {Cohort 4};
\node[font={\normalsize}] at (3,7) {SP1};
\node[font={\normalsize}] at (23,7) {SP2};
\node[font={\normalsize}] at (43,7) {SP3};
\node[font={\normalsize}] at (63,7) {SP4};
\node[font={\normalsize}] at (83,7) {SP5};
\node[font={\normalsize}] at (103,7) {SP6};
\node[font={\normalsize}] at (3,-62) {$\mathcal{P}_1$};
\node[font={\normalsize}] at (23,-62) {$\mathcal{P}_2$};
\node[font={\normalsize}] at (43,-62) {$\mathcal{P}_3$};
\node[font={\normalsize}] at (83,-62) {$\mathcal{P}_4$};
\node (image) at (0,-68) {\includegraphics[width=11.35cm]{cor-T_colourkey_horiz.pdf}};
\end{tikzpicture}
\end{figure}

Consider $K$ independent cohorts with respective sample sizes $n_k$, $k=1, \ldots, K$. In each cohort we observe $J$ correlated $m_{i,j}$-element vector outcomes $\by_{i,jk}=(y_{i1,jk}, \ldots, y_{im_{i,j},jk})^T$, $j=1, \ldots, J$, for each participant $i$, $i=1, \ldots, n_k$, with $\bx_{i,jk}$ the corresponding $m_{i,j} \times p$ covariate matrix. Here $\bx_{i,jk}$ is assumed to be the cohort- and outcome-specific observations on the same variables across outcomes and cohorts (e.g. age, sex, exposure). Participants are assumed independent, and let $\bSigma_{i,k}$ be the covariance matrix of $\by_{i,k}=(\by_{i,1k}, \ldots, \by_{i,Jk})^T$. We consider the model $E(\by_{ir,jk})=h_{jk}(\bx_{ir,jk} \btheta_{jk})$, $r=1, \ldots, m_{i,j}$, where $h_{jk}$ is a known link function and $\btheta_{jk}$ is a $p\times 1$ parameter vector of interest. Suppose there exists a known partition $\mathcal{P}=\{ \mathcal{P}_g \}_{g=1}^G$, $\mathcal{P}$ a set of disjoint non-empty subsets $\mathcal{P}_g$, of $\{(j,k)\}_{j,k=1}^{J,K}$ such that $\btheta_{jk}\equiv \btheta_g$ and $h_{jk} \equiv h_g$ for $(j,k) \in \mathcal{P}_g$. There are $G$ unique values of $\btheta_{jk}$, $j=1, \ldots, J$, $k=1, \ldots, K$. Let $\mathcal{P}_g$ have cardinality $d_g$ such that $\sum_{g=1}^G d_g =JK$. We want to estimate and make inference about the true value $\btheta_0=(\btheta_{0,g})_{g=1}^G \in \mathbb{R}^{Gp}$ of $\btheta=(\btheta_g)_{g=1}^G \in \mathbb{R}^{Gp}$ based on all $JK$ sources of information.\\
We give an example from Section \ref{sec:data} to fix ideas. For $K=4$ cohorts, we quantify 36 metabolites from $J=6$ peptide sub-pathways: acetylated peptides, dipeptide, fibrinogen cleavage peptide, gamma-glutamyl amino acid, modified peptides, and polypeptide. Given the biological function of these sub-pathways, we model the effect of smoking on the peptide metabolites by integrating its effect over the four cohorts and the latter four sub-pathways, and integrating the effect of smoking on the first, second and third sub-pathways only over cohorts. This partition corresponds to $\mathcal{P}=\{\mathcal{P}_g \}_{g=1}^4$, where $\mathcal{P}_1=\{ (1,k)\}_{k=1}^K$, $\mathcal{P}_2=\{(2,k)\}_{k=1}^K$, $\mathcal{P}_3=\{ (3,k)\}_{k=1}^K$ and $\mathcal{P}_4=\{(4,k),(5,k),(6,k) \}_{k=1}^K$. This partition is visualized in Figure \ref{peptide-partition}.\\
The proposed method creates a set of moment conditions on $\btheta$, with corresponding estimators, from each data source. We propose an efficient and computationally attractive estimator that linearly updates data source-specific estimators by weighting them as a function of their covariance.\\
We introduce some notation to facilitate the description of the proposed method in sections \ref{subsec:QIF} and \ref{subsec:integration}. For ease of exposition, we henceforth use the term ``cohorts'' to refer to the $K$ disjoint and independent participant groups, ``block'' to refer to the $J$ correlated vector outcomes. We refer to cohort $k$ and block $j$ as data source $(j,k)$. Let $\| \cdot \|$ denote the $L_2$ norm on vectors and the Frobenius norm on matrices. Define the stacking operators $(\cdot)_{(j,k)\in \mathcal{P}_g}$ and $(\cdot)_{g=1}^G$ for vectors $\{ \ba_{jk} \}_{j=1,k=1}^{J,K}$, $\ba_{jk} \in \mathbb{R}^{D_{jk}}$ and matrices $\{ \bA_{jk} \}_{j=1,k=1}$, $\bA_{jk} \in \mathbb{R}^{D_{1jk}\times D_2}$, as 
\begin{align*}
\left( \ba_{jk} \right)_{(j,k)\in \mathcal{P}_g}&=\ba_g=\left( \begin{array}{ccc} \ba^T_{j_1k_1} & \ldots & \ba^T_{j_{d_g} k_{d_g}} \end{array} \right)^T , \quad \mathcal{P}_g=\left\{(j_1,k_1), \ldots, (j_{d_g}, k_{d_g}) \right\},\\
\left( \bA_{jk} \right)_{(j,k)\in \mathcal{P}_g}&=\bA_g=\left( \begin{array}{ccc} \bA^T_{j_1k_1} & \ldots & \bA^T_{j_{d_g} k_{d_g}} \end{array} \right)^T , \quad \mathcal{P}_g=\left\{(j_1,k_1), \ldots, (j_{d_g}, k_{d_g}) \right\},\\
\left( \ba_g \right)_{g=1}^G &=\left( \begin{array}{ccc} \ba^T_1 & \ldots & \ba^T_G \end{array} \right)^T ,\quad \left( \bA_g \right)_{g=1}^G=\left( \begin{array}{ccc} \bA^T_1 & \ldots & \bA^T_G \end{array} \right)^T .
\end{align*}
Let $\ba^{\otimes 2}$ denote the outer product of a vector $\ba$ with itself, namely $\ba^{\otimes 2}=\ba \ba^T$. Denote $N=\sum_{k=1}^K n_k$ the combined sample size over $K$ cohorts. For each participant $i$, denote the combined $M_i$-dimensional response $\by_i=\left(\by_{i,1}, \ldots, \by_{i,J} \right)^T$ over the $J$ blocks such that $\sum_{j=1}^{J} m_{i,j} = M_i$ for each $i=1, \ldots, N$. Combination across data sources is not restricted to the order of data entry: responses may be grouped according to pre-specified data source memberships, according to, say, substantive scientific knowledge. In this paper, with no loss of generality, we use the order of data entry in the data combination procedure.\\
Our proposed method can also be applied as a divide-and-conquer procedure to a large dataset with $N$ samples on $M$ correlated outcomes. Dividing this large dataset into $JK$ sources of data with sample size $n_k$ and $m_j$-dimensional outcomes yields the above framework with the simplification $M_i=M$, $m_{i,j}=m_j$.

\subsection{Quadratic inference functions}
\label{subsec:QIF}
We propose to first obtain \cite{Qu-Lindsay-Li}'s quadratic inference function estimator of $\btheta_{jk}$ in data source $(j,k)$. This is a standard analysis that is performed on each data source individually as if there was no other source of data to improve estimation.
Consider an arbitrary data source $(j,k)$. Let $\bmu_{i,jk}=E(\by_{i,jk})$ the $m_{i,j}$-dimensional mean of the outcome $\by_{i,jk}$ for $i=1, \ldots, n_k$. Let $\dot{\bmu}^{\btheta}_{i,jk}=\partial \bmu_{i,jk}/\partial \btheta_{jk}$ be an $m_{i,j} \times p$-dimensional partial derivative matrix. Following \cite{Qu-Lindsay-Li}, we approximate the inverse working correlation matrix of $\by_{i,jk}$ by $\sum_{s=1}^{s_{jk}} b_{s,jk} \bB_{s,jk}$ where $b_{1,jk}, \ldots, b_{s_{jk},jk}$ are unknown constants and $\bB_{1,jk}, \ldots, \bB_{s_{jk},jk}$ are known basis matrices with elements $0$ and $1$.  Let
\begin{align}
&\bPsi_{jk}(\btheta_{jk})=
\frac{1}{n_k} \sum \limits_{i=1}^{n_k} \bpsi_{i,jk}(\btheta_{jk})=\frac{1}{n_k} \sum \limits_{i=1}^{n_k} 
\left( \begin{array}{c} 
\dot{\bmu}^{\btheta~T}_{i,jk} \bD^{-\frac{1}{2}}_{i,jk} \bB_{1,jk} \bD^{-\frac{1}{2}}_{i,jk} (\by_{i,jk}-\bmu_{i,jk})\\
\vdots \\
\dot{\bmu}^{\btheta~T}_{i,jk} \bD^{-\frac{1}{2}}_{i,jk} \bB_{s_{jk},jk} \bD^{-\frac{1}{2}}_{i,jk} (\by_{i,jk}-\bmu_{i,jk})
\end{array} \right),\label{block-EE-1}
\end{align}
where $\bD_{i,jk}$ is the diagonal marginal covariance matrix of $\by_{i,jk}$, and $s_{jk}$ is typically chosen as $s_{jk}=2$.  Let $\bC_{jk}=(1/n_k) \sum_{i=1}^{n_k} \bpsi^{\otimes 2}_{i,jk}(\btheta_{jk})$, which depends only on $\btheta_{jk}$. The quadratic inference function takes the form $Q_{jk}(\btheta_{jk})=n_k \bPsi^T_{jk}(\btheta_{jk}) \bC^{-1}_{jk} \bPsi_{jk}(\btheta_{jk})$, and the data source-specific quadratic inference function estimator is $\widehat{\btheta}_{jk}=\arg \min_{\btheta_{jk}} Q_{jk}(\btheta_{jk})$. No nuisance correlation parameter is involved in the estimation. Under mild regularity conditions, $\widehat{\btheta}_{jk}$ is consistent and asymptotically normal \citep{Hansen}. When the working correlation structure is correctly specified by the basis matrix expansion, this estimator is semi-parametrically efficient, i.e. as efficient as the quasilikelihood; even when the working correlation structure is misspecified, this estimator is still efficient within a general family of estimators \citep{Qu-Lindsay-Li}. These advantageous properties allow us to derive an efficient integrated estimator in section \ref{subsec:integration}. 

\subsection{Integrated estimator}
\label{subsec:integration}
Define the cohort indicator $\delta_i(k)=\mathbbm{1}(\mbox{participant }i \allowbreak \mbox{ is in cohort }k)$ for $i=1, \ldots, N$, $k=1, \ldots, K$. For participant $i$, let 
\begin{align*}
\bpsi_{i,g}(\btheta)&=\left\{ \delta_i(k) \bpsi_{i,jk}(\btheta_{jk}) \right\}_{(j,k)\in \mathcal{P}_g},\quad
\bpsi_i(\btheta)= \left\{ \bpsi_{i,g}(\btheta) \right\}_{g=1}^G.
\end{align*}
Then we can define $\bPsi_N(\btheta)=(1/N) \sum_{i=1}^N \bpsi_i(\btheta) $. It is easy to show that
\begin{align*}
\bPsi_N(\btheta)&=\frac{1}{N} \sum \limits_{i=1}^N \left[ \left\{ \delta_i(k) \bpsi_{i,jk}(\btheta_{jk}) \right\}_{(j,k) \in \mathcal{P}_g} \right]_{g=1}^G=\frac{1}{N} \left[ \left\{ n_k \bPsi_{jk}(\btheta_{jk}) \right\}_{(j,k)\in \mathcal{P}_g} \right]_{g=1}^G.
\end{align*}
We define a few sample sensitivity matrices. For data source $(j,k)$, define the $(ps_{jk})\times p$-dimensional sample sensitivity matrix $\widehat{\bS}_{jk}=-\{ \nabla_{\btheta_{jk}} \bPsi_{jk}(\btheta_{jk}) \} \lvert_{\btheta_{jk}=\widehat{\btheta}_{jk}}$. For the $g$th set $\mathcal{P}_g$, define $\widehat{\bS}_g=(n_k \widehat{\bS}_{jk})_{(j,k)\in \mathcal{P}_g}$ the matrix of stacked sensitivity matrices with row-dimension $\sum_{(j,k) \in \mathcal{P}_g} ps_{jk}$ and column-dimension $p$. Finally, let $\widehat{\bS}=\mbox{blockdiag} \{ \widehat{\bS}_g \}_{g=1}^G$ the sample sensitivity matrix of $\bPsi_N$ with row-dimension $\sum_{g=1}^G \sum_{(j,k) \in \mathcal{P}_g} ps_{jk} = \sum_{j,k=1}^{J,K} ps_{jk}$ and column-dimension $Gp$.\\
Let $\widehat{\btheta}_g=(\widehat{\btheta}_{jk})_{(j,k)\in \mathcal{P}_g}$ and $\widehat{\btheta}_{list}= ( \widehat{\btheta}_g )_{g=1}^G$. Define $\bpsi_i(\widehat{\btheta}_{list})=[ \{ \delta_i(k) \bpsi_{i,jk}(\widehat{\btheta}_{jk}) \}_{(j,k)\in \mathcal{P}_g} ]_{g=1}^G$. Let $\widehat{\bV}_N=(1/N) \sum_{i=1}^N \{ \bpsi_i(\widehat{\btheta}_{list}) \}^{\otimes 2}$ be the sample covariance of $\bPsi_N(\btheta_0)$ with row- and column-dimension $\sum_{j,k=1}^{J,K} ps_{jk}$. Then we define the integrated estimator of $\btheta$ as
\begin{align}
\widehat{\btheta}&=\left( \widehat{\bS}^T \widehat{\bV}^{-1}_N \widehat{\bS} \right)^{-1} \widehat{\bS}^T \widehat{\bV}^{-1}_N  \left\{ (n_k \widehat{\bS}_{jk} \widehat{\btheta}_{jk} )_{(j,k)\in \mathcal{P}_g} \right\}_{g=1}^G.
\label{def:one-step-estimator-both}
\end{align}
Following similar steps to \cite{Hector-Song-2020}, we can show this integrated estimator is asymptotically equivalent to the minimizer of the optimal combination of the moment conditions. Estimators from different sets $\mathcal{P}_g$ may not be combined but still benefit from correlation between data sources, captured by $\widehat{\bV}_N$, leading to improved statistical efficiency. This is similar to the gain in efficiency in seemingly unrelated regression \citep{Zellner}. The closed-form estimator in \eqref{def:one-step-estimator-both} depends only on estimators, estimating equations and sample sensitivity matrices from each data source. It can be implemented in a fully parallelized MapReduce framework, where data sources are analyzed in parallel on distributed nodes using quadratic inference functions and results from the separate analyses are sent to a main node to compute the integrated estimator. This procedure is privacy-preserving, since the combination step does not require access to individual level data, and communication-efficient, since it does not require multiple rounds of communication between the main and distributed nodes. In addition, it is computationally efficient at each node since nuisance correlation parameters are not involved in the estimation.\\
Two special cases of interest arise when $\mathcal{P}_g$ are all singletons ($G=JK$, $d_g=1$ for $g=1, \ldots, G$) and when $\mathcal{P}=\{(j,k) \}_{j,k=1}^{J,K}$ ($G=1$, $d_1=JK$). The former case reduces to seemingly unrelated regression, in which $JK$ regression equations are used to estimate $JK$ parameter vectors. The latter case corresponds to a fully integrated analysis of all $JK$ data sources similar in spirit to meta-analysis. The estimator $\widehat{\btheta}$ in \eqref{def:one-step-estimator-both} takes a special form: let $\widehat{\bV}_N=\mbox{blockdiag} \{ (n_k/N) \widehat{\bV}_k \}_{k=1}^K$ with block matrices
\begin{align*}
\widehat{\bV}_k=\frac{1}{n_k} \sum \limits_{i=1}^{n_k} \left\{ \bpsi_{i,k}(\widehat{\btheta}_k) \right\}^{\otimes 2} \quad (k=1, \ldots, K).
\end{align*}
Let $[ \widehat{\bV}_{k}]^{i;j}$ denote the rows and columns of $\widehat{\bV}^{-1}_k$ corresponding to blocks $i$ and $j$ respectively and define the sample Godambe information $\widehat{\bJ}_{ijk}= \widehat{\bS}_{ik} \left[ \widehat{\bV}_{k} \right]^{i;j} \widehat{\bS}_{jk}$ \citep{Godambe-Heyde, Song}. The integrated estimator simplifies to
\begin{align*}
\widehat{\btheta}&=\left( \sum \limits_{k=1}^K \sum \limits_{i,j=1}^J n_k\widehat{\bJ}_{ijk} \right)^{-1} \sum \limits_{k=1}^K \sum \limits_{i,j=1}^J n_k\widehat{\bJ}_{ijk} \widehat{\btheta}_{jk}.
\end{align*}

Inversion of $\widehat{\bV}_N$ may be numerically unstable or undefined in some settings. When $J$, $K$ and/or $p$ are large, the large dimension of $\widehat{\bV}_N$ can lead to numerical difficulties in its inversion. Using an equicorrelated structure for the data source analysis can also lead to a rank-deficient weight matrix $\widehat{\bV}_N$ \citep{Hu-Song}. To handle these cases we propose to reduce the number of estimating equations similarly to \cite{Cho-Qu}: principal components of $\bPsi_N$ with non-zero eigenvalues are selected so as to maximize the variability explained and eliminate between-component correlations. These linear combinations of the original estimating equations have lower dimension than $\bPsi_N$ and yield an invertible sample variability matrix $\widehat{\bV}_N$. The method described in Section \ref{sec:method} remains unchanged with the substitution of the principal components for $\bPsi_N$.

\subsection{Large sample theory}
\label{subsec:theory}

Let $n_{\min}=\min_{k=1, \ldots, K} n_k$. Define the sensitivity matrices $\bs_{jk}(\btheta_{jk})=-\nabla_{\btheta_{jk}} E_{\btheta_{0,g}} \{ \bpsi_{i,jk} (\btheta_{jk}) \}$ for $(j,k) \in \mathcal{P}_g$, $s_g(\btheta_g)= \{ (n_k/N) \bs_{jk}(\btheta_{jk})\}_{(j,k) \in \mathcal{P}_g}$, and $\bs(\btheta)=\mbox{blockdiag} \{ \bs_g(\btheta_g) \}_{g=1}^G$. Define the variability matrix $\bv(\btheta)=Var_{\btheta_0} \{\bpsi_i(\btheta) \}$. Regularity conditions required to establish the consistency and asymptotic normality of the integrated estimator $\widehat{\btheta}$ in \eqref{def:one-step-estimator-both} are listed in Appendix \ref{appendix:conditions}. In particular, assumption \ref{QIF-conds} guarantees the consistency and asymptotic normality of the data source-specific estimators $\widehat{\btheta}_{jk}$, and assumption \ref{DDIMM-conds} guarantees the consistency and asymptotic normality of the integrated estimator $\widehat{\btheta}$ in \eqref{def:one-step-estimator-both}. These results are summarized in Theorem \ref{thm:norm}.

\begin{theorem}[Consistency and asymptotic normality]
\label{thm:norm}
Suppose assumptions \ref{QIF-conds}-\ref{DDIMM-conds} hold. Let $\bj(\btheta)= \lim_{n_{\min} \rightarrow \infty} \bs^T(\btheta) \bv^{-1}(\btheta) \bs(\btheta)$ denote the Godambe information matrix of $\bPsi_N$. As $n_{\min} \rightarrow \infty$, $\sqrt{N} ( \widehat{\btheta}-\btheta_0 ) \stackrel{d}{\rightarrow} \mathcal{N} (0, \bj^{-1}(\btheta_0) )$, and $\bj(\btheta)$ has $(r,t)$th block element $\lim_{n_{\min} \rightarrow \infty} \{ \bs^T_r(\btheta_0) [\bv(\btheta_0)]^{r;t} \bs_t(\btheta_0) \}$ where $[\bv(\btheta_0)]^{r;t}$ is the submatrix of $\bv^{-1}(\btheta_0)$ consisting of rows and columns corresponding to partitions $r$ and $t$ respectively, $r,t=1, \ldots, G$.
\end{theorem}
The proof of Theorem \ref{thm:norm} can be done similarly to Theorem 9 in \cite{Hector-Song-JMLR} and is omitted. It is clear from Theorem \ref{thm:norm} that the asymptotic covariance of $\widehat{\btheta}$ can be consistently estimated by the sandwich covariance $N (\widehat{\bS}^T \widehat{\bV}^{-1}_N \widehat{\bS} )^{-1}$. A goodness-of-fit test is available from Theorem \ref{thm:test} below to check the validity of modelling assumptions and appropriateness of the data source partition $\mathcal{P}$.

\begin{theorem}[Homogeneity test]
\label{thm:test}
Suppose assumptions \ref{QIF-conds}-\ref{DDIMM-conds} hold with $\widehat{\btheta}$ defined in \eqref{def:one-step-estimator-both}. Then as $n_{\min} \rightarrow \infty$, the statistic $Q_N(\widehat{\btheta})=N\bPsi^T_N(\widehat{\btheta}) \widehat{\bV}^{-1}_N \bPsi_N(\widehat{\btheta})$ converges in distribution to a $\chi^2$ random variable with degrees of freedom $\sum_{j,k=1}^{J,K} ps_{jk} -Gp$.
\end{theorem}
The proof of Theorem \ref{thm:test} follows from \cite{Hansen} and \cite{Hector-Song-JMLR}. In practice, the computation of the quadratic test statistic in Theorem \ref{thm:test} can be implemented in a distributed fashion despite requiring access to individual data sources to recompute $\bPsi_N(\widehat{\btheta})$. Theorem \ref{thm:test} is particularly useful to compare the fit from different data source partitions, and can be used to detect inappropriate modelling and strong data heterogeneity requiring modification of the integration partition. Let $\mathcal{P}^i=\{ \mathcal{P}^i_g \}_{g=1}^{G^i}$ and $\mathcal{P}^h=\{ \mathcal{P}^h_g \}_{g=1}^{G^h}$ two data source partitions such that $\mathcal{P}^i$ is itself a nested partition of $\mathcal{P}^h$; let $Q^i_N(\widehat{\btheta}^i)$ and $Q^h_N(\widehat{\btheta}^h)$ be the statistics from Theorem \ref{thm:test} based on partitions $\mathcal{P}^i$ and $\mathcal{P}^h$ respectively, where the same working correlation structures and mean models are used for both. Then a test statistic of the null hypothesis of parameter homogeneity in partition $\mathcal{P}^i$, $H_0: \btheta_{jk}=\btheta^i_g$ for all $(j,k) \in \mathcal{P}^i_g$, $g=1, \ldots, G^i$, can be formulated as
\begin{align}
\label{Q-statistic}
Q=Q^i_N(\widehat{\btheta}^i)-Q^h_N(\widehat{\btheta}^h), 
\end{align}
which under $H_0$ is asymptotically $\chi^2$ distributed with degrees of freedom $(G^h-G^i)p$. Failure to reject the null hypothesis implies the smaller partition $\mathcal{P}^i$ fits as well or better than the larger partition $\mathcal{P}^h$. Asymptotically, the power of this test goes to $1$ as $N \rightarrow \infty$. In practice, as $N$ grows large, this test has the power to detect small deviations from the null hypothesis and therefore will reject the null hypothesis, which may not be practically useful. Alternatively, partitions can be compared by minimizing the GMM-BIC of \cite{Andrews},
\begin{align}
\label{GMM-BIC}
BIC(\mathcal{P}) = Q_N(\widehat{\btheta}) -\log(N) \left( \sum_{j,k=1}^{J,K} ps_{jk} -Gp \right),
\end{align}
for some partition $\mathcal{P}$ with estimator $\widehat{\btheta}$.\\
Lastly, we discuss estimation efficiency of our proposed integrated estimator $\widehat{\btheta}$ in \eqref{def:one-step-estimator-both}, which is asymptotically equivalent to \cite{Hansen}'s optimal generalised method of moments estimator $\widehat{\btheta}_{opt}=
\arg \min_{\btheta} 
\bPsi^T_N(\btheta)
\widehat{\bV}^{-1}_N 
\bPsi_N(\btheta)$. The optimality of this estimator is achieved within the class of estimators minimizing the quadratic form $\bPsi^T_N(\btheta) \bW \bPsi_N(\btheta)$ with positive semi-definite matrices $\bW$ \citep{Hansen}. Additionally, in Theorem \ref{thm:efficiency} we show the efficiency gain from combining estimators over data sources for an arbitrary data source $(j,k) \in \mathcal{P}_g$. The asymptotic covariance of $\widehat{\btheta}_{jk}$ is larger than or equal to (in the L\"{o}wner partial ordering) the asymptotic covariance of the subvector of $\widehat{\btheta}$ corresponding to $\mathcal{P}_g$, $\widehat{\btheta}^g$.

\begin{theorem}[Efficiency gain]
\label{thm:efficiency}
Suppose assumptions \ref{QIF-conds}-\ref{DDIMM-conds} hold with $\widehat{\btheta}$ defined in \eqref{def:one-step-estimator-both}. Consider an arbitrary data source $(j,k) \in \mathcal{P}_g$ for some $g \in \{1, \ldots, G \}$.   The asymptotic covariances, denoted by $Avar$, of $\widehat{\btheta}_{jk}$ and $\widehat{\btheta}^g$ satisfy $Avar(\sqrt{N} \widehat{\btheta}^g) \preceq \{ \lim_{n_k \rightarrow \infty} (N/n_k) \} Avar(\sqrt{n_k} \widehat{\btheta}_{jk})$, where $\preceq$ denotes L\"{o}wner's partial ordering in the space of nonnegative definite matrices.
\end{theorem}
The proof of Theorem \ref{thm:efficiency} is given in Appendix \ref{appendix:proofs}. The gain in statistical efficiency given by Theorem \ref{thm:efficiency} is due to the use of between-data source correlation, captured by $\widehat{\bV}_N$, and to the combination of estimators within each $\mathcal{P}_g$. \\
Finally, we remark that the proposed method is a generalization of \cite{Wang-Wang-Song-2012}, which only allows for combining over independent data sources. Here we introduce a non-diagonal weight matrix $\widehat{\bV}_N$ to incorporate correlation between data sources, leading to improved statistical efficiency. We also propose a closed form integrated estimator that is more computationally advantageous than their iterative minimization procedure, leading to improved computational scalability.

\section{Simulations}
\label{sec:simulations}

We examine the performance of the integrated estimator $\widehat{\btheta}$ through three sets of simulations. In the first and third sets, for simplicity $\mathcal{P}=\{(j,k)\}_{j,k=1}^{J,K}$ ($G=1$, $d_1=JK$) and $M_i=M$ for $i=1, \ldots, N$. The second set explores the performance of the selective combination scheme with a partition of $\{(j,k)\}_{j,k=1}^{J,K}$ and confirms the distribution of statistic $Q_N(\widehat{\btheta})$ in Theorem \ref{thm:test}. Simulations are conducted using R software on a standard Linux cluster. In all simulations, covariates consist of an intercept and two independent $M$-dimensional continuous variables drawn from Multivariate Normal distributions with non-diagonal covariance matrices. True values of the regression parameters are drawn from uniform distributions on $(-5,5)$.\\
The first set of simulations considers the logistic regression $\log\{ \mu_{ir,jk}/(1-\mu_{ir,jk}) \}=\bX_{ir,jk} \btheta$ with $\mu_{ir,jk}=E(Y_{ir,jk} \lvert \bX_{ir,jk}, \btheta)$, $r=1, \ldots, m_j$, where $\bY_i$ is a $M$-variate correlated Bernoulli random variable. We illustrate the finite sample performance of $\widehat{\btheta}$ in two settings: in Setting I, $K=2$ with $n_1=n_2=5000$, $J=4$ with block response dimensions $163$, $181$, $260$, $396$ such that $M=1000$; in Setting II, $K=4$ with $n_1=n_2=n_3=n_4=5000$, $J=8$ with block response dimensions $227$, $252$, $357$, $381$, $368$, $276$, $226$, $413$ such that $M=2500$. The true value of $\btheta$ is set to $\btheta_0=(-4.44, 1.11, -2.22)$. $\bY_i$ is simulated using the \verb|SimCorMultRes| R package \citep{Touloumis} with data source-specific AR(1) correlation structures. We estimate $\btheta$ with an AR(1) working data source correlation structure ($s_{jk}=2$, $j=1, \ldots, J$, $k=1, \ldots, K$). Root mean squared error (RMSE), empirical standard error (ESE), asymptotic standard error (ASE), mean bias (B), 95\% confidence interval coverage (CI), 95\% confidence interval length (L) and type-I error (ERR) of $\widehat{\btheta}$ averaged over 500 simulations are presented in Table \ref{simulations-logistic}. We see from Table \ref{simulations-logistic} that the ASE of $\widehat{\btheta}$ approximates the ESE, confirming the covariance formula in Theorem \ref{thm:norm}. Additionally, $\widehat{\btheta}$ appears consistent since RMSE, ASE and ESE are approximately equal, and the bias B is negligible. We observe appropriate 95\% confidence interval coverage and proper Type-I error control. The mean CPU times for this analysis are 1.4 and 2.1 minutes for Settings I and II respectively. By comparison, an analysis using quadratic inference functions on the entire combined data has mean CPU times of 10 minutes and 2.4 hours for Settings I and II respectively, highlighting the considerable computational advantage of our proposed method. Details and simulation metrics for this comparative analysis averaged over 500 simulations are available in the Supplemental Material \citep{Hector-Song-DIQIF}. Confidence interval lengths are shorter for our method than for the quadratic inference function over the entire combined data under the same 95\% coverage, highlighting our method's gain in statistical efficiency obtained from better modeling of the correlation structure of the outcome. We attempted to compare our method to an analysis using generalized estimating equations (using R package \verb|geepack|) on the entire combined data but the mean CPU time exceeded 63 hours in Setting I using the same Linux cluster as in previous simulations, which we view as too computationally burdensome, and we do not report results.
\begin{table}[ht]
\centering
\caption{Logistic regression simulation results with homogeneity partition $\mathcal{P}=\{(j,k)\}_{j,k=1}^{J,K}$ ($G=1$, $d_1=JK$). \label{simulations-logistic}}
\begin{subtable}[ht]{\textwidth}
\centering
\caption{Setting I: $K=2$ with $n_1=n_2=5000$, $J=4$ with block response dimensions $163$, $181$, $260$, $396$ such that $M=1000$. \label{simulations-logistic-1}}
\begin{tabular}{lrrrrrrr}
 & RMSE$\times 10^{-3}$ & ESE$\times 10^{-3}$ & ASE$\times 10^{-3}$ & B$\times 10^{-4}$ & CI & L$\times 10^{-3}$ & ERR \\ 
  Intercept & 4.89 & 4.86 & 4.83 & $-5.84$ & 0.95 & 18.74 & 0.05 \\ 
  $\bX_1$ & 1.43 & 1.42 & 1.40 & 1.79 & 0.94 & 5.45 & 0.06 \\ 
  $\bX_2$ & 2.45 & 2.43 & 2.48 & $-3.49$ & 0.95 & 9.65 & 0.05 \\ 
\end{tabular}
\end{subtable}
\hfill
\begin{subtable}[ht]{\textwidth}
\centering
\caption{Setting II: $K=4$ with $n_1=n_2=n_3=n_4=5000$, $J=8$ with block response dimensions $227$, $252$, $357$, $381$, $368$, $276$, $226$, $413$ such that $M=2500$. \label{simulations-logistic-2}}
\begin{tabular}{lrrrrrrr}
 & RMSE$\times 10^{-3}$ & ESE$\times 10^{-3}$ & ASE$\times 10^{-3}$ & B$\times 10^{-4}$ & CI & L$\times 10^{-3}$ & ERR \\ 
Intercept & 2.29 & 2.20 & 2.21 & $-6.31$ & 0.95 & 8.62 & 0.05 \\ 
  $\bX_1$ & 0.66 & 0.64 & 0.63 & 1.76 & 0.95 & 2.48 & 0.05 \\ 
  $\bX_2$ & 1.19 & 1.14 & 1.13 & $-3.36$ & 0.95 & 4.42 & 0.05 \\ 
\end{tabular}
\end{subtable}
\end{table}
\newline The second set of simulations again considers the logistic regression $\log(\mu_{i,jk}/(1-\mu_{i,jk}))=\bX_{i,jk} \btheta$ with $\bmu_{i,jk}=E(\bY_{i,jk} \lvert \bX_{i,jk}, \btheta)$, where $\bY_i$ is a $M$-variate correlated Bernoulli random variable of dimension $M=500$ from $J=5$ blocks with $(m_1, \ldots, m_5)=(130,75,92,115,88)$. We consider the integration of two cohorts of respective sample sizes $n_1=n_2=5000$. The underlying partition is $\mathcal{P}=\{\mathcal{P}_g\}_{g=1}^G$ with $G=3$, $\mathcal{P}_1=\{(1,k),(2,k)\}_{k=1}^K$, $\mathcal{P}_2=\{(3,k)\}_{k=1}^K$ and $\mathcal{P}_3=\{(4,k),(5,k)\}_{k=1}^K$ with respective true values $\btheta_{0,1}=(-4.44, 1.11, -2.22)$, $\btheta_{0,2}=(0.222,-0.888,-0.444)$ and $\btheta_{0,3}=(-1.554,\allowbreak -3.108,0.777)$. $\bY_i$ is simulated as in the first set of simulations. We estimate $\btheta$ and present summary results averaged over 500 simulations in Table \ref{simulations-logistic-selective-CS} for the exchangeable working data source correlation structure ($s_{jk}=2$, $j=1, \ldots, J$, $k=1, \ldots, K$) and in the Supplementary Material \citep{Hector-Song-DIQIF} for the AR(1) working data source correlation structure ($s_{jk}=2$, $j=1, \ldots, J$, $k=1, \ldots, K$). From Table \ref{simulations-logistic-selective-CS} we again observe correct estimation of the asymptotic covariance, minimal bias and proper Type-I error control for regression parameters. The integrative procedure seems to work well with partial heterogeneity of mean effects. The mean CPU time for this analysis is $11.2$ seconds.
\begin{table}[ht]
\centering
\caption{Logistic regression simulation results with $\mathcal{P}=\{\mathcal{P}_1, \mathcal{P}_2, \mathcal{P}_3 \}$, $\mathcal{P}_1=\{(1,k),(2,k)\}_{k=1}^K$, $\mathcal{P}_2=\{(3,k)\}_{k=1}^K$ and $\mathcal{P}_3=\{(4,k),(5,k)\}_{k=1}^K$, and exchangeable working data source correlation structure. \label{simulations-logistic-selective-CS}}
\begin{subtable}[ht]{\textwidth}
\centering
\caption{Summary results for $\mathcal{P}_1=\{(1,k),(2,k)\}_{k=1}^K$. \label{simulations-logistic-selective-1}}
\begin{tabular}{lrrrrrrr}
 & RMSE$\times 10^{-3}$ & ESE$\times 10^{-3}$ & ASE$\times 10^{-3}$ & B$\times 10^{-4}$ & CI & L$\times 10^{-3}$ & ERR \\ 
Intercept & 10.89 & 10.62 & 10.30 & $-24.48$ & 0.93 & 40.06 & 0.07 \\ 
  $\bX_1$ & 3.16 & 3.11 & 3.03 & 5.94 & 0.94 & 11.84 & 0.06 \\ 
  $\bX_2$ & 5.58 & 5.44 & 5.36 & $-12.26$ & 0.93 & 20.86 & 0.07 \\ 
\end{tabular}
\end{subtable}
\hfill
\begin{subtable}[ht]{\textwidth}
\centering
\caption{Estimates for $\mathcal{P}_2=\{(3,k)\}_{k=1}^K$. \label{simulations-logistic-selective-2}}
\begin{tabular}{lrrrrrrr}
 & RMSE$\times 10^{-3}$ & ESE$\times 10^{-3}$ & ASE$\times 10^{-3}$ & B$\times 10^{-4}$ & CI & L$\times 10^{-3}$ & ERR \\ 
Intercept & 3.22 & 3.22 & 3.37 & $-2.16$ & 0.95 & 12.93 & 0.05 \\ 
  $\bX_1$ & 2.25 & 2.25 & 2.21 & $-0.48$ & 0.95 & 8.57 & 0.05 \\ 
  $\bX_2$ & 1.51 & 1.51 & 1.53 & $-0.06$ & 0.95 & 5.98 & 0.05 \\ 
  \end{tabular}
\end{subtable}
\hfill
\begin{subtable}[ht]{\textwidth}
\centering
\caption{Estimates for $\mathcal{P}_3=\{(4,k),(5,k)\}_{k=1}^K$. \label{simulations-logistic-selective-3}}
\begin{tabular}{lrrrrrrr}
 & RMSE$\times 10^{-3}$ & ESE$\times 10^{-3}$ & ASE$\times 10^{-3}$ & B$\times 10^{-4}$ & CI & L$\times 10^{-3}$ & ERR \\ 
 Intercept & 4.72 & 4.69 & 4.80 & $-6.06$ & 0.95 & 18.66 & 0.05 \\ 
  $\bX_1$ & 6.36 & 6.25 & 6.31 & $-11.73$ & 0.94 & 24.48 & 0.06 \\ 
  $\bX_2$ & 2.03 & 2.01 & 2.04 & 2.82 & 0.95 & 7.98 & 0.05 \\ 
  \end{tabular}
\end{subtable}
\end{table}
\newline The third set of simulations considers the linear regression $\bmu_{i,jk}=\bX_{i,jk} \btheta$ with $\bmu_{i,jk}=E(\bY_{i,jk} \lvert \bX_{i,jk}, \btheta)$, where $\bY_i \sim \mathcal{N}(\bX_i \btheta, \bSigma)$. We illustrate the finite sample performance of $\widehat{\btheta}$ with $K=10$ cohorts where $n_k=10000$ for all $k$ for a total sample size of $N=100000$, and $J=250$ response blocks where $m_j=400$ for all $j$ for a total response dimension $M=100000$ of $\bY$. The true value of $\btheta$ is set to $\btheta_0=(1.1, 2.2, 3.3)^T$. Responses are simulated from a Multivariate Normal distribution with block-AR(1) covariance structure with data source-specific variance and correlation parameters. We estimate $\btheta$ with an AR(1) working data source correlation structure ($s_{jk}=2$, $j=1, \ldots, J$, $k=1, \ldots, K$). RMSE, ESE, ASE, B, CI, L and ERR of $\widehat{\btheta}$ averaged over 500 simulations are presented in Table \ref{simulations-huge}. We observe in Table \ref{simulations-huge} slight inflation of Type-I error due to under-estimation of the asymptotic covariance. This is potentially due to the high-dimensionality of $\bPsi_N$ and $\widehat{\bV}_N$, which have dimension $15000$, leading to numerical instability. This under-estimation is similar to the generalised method of moments case and is discussed in Section \ref{sec:discussion}. The performance of our method in this ultra-high dimension is nonetheless remarkable: with $10^{10}$ data points with high-variability in both outcomes and covariates, the procedure is able to estimate and infer the true mean effects with minimal bias and only slight under-coverage. The mean CPU time for this analysis is 17.9 hours.
\begin{table}[ht]
\centering
\caption{Linear regression simulation results with homogeneity partition $\mathcal{P}=\{(j,k)\}_{j,k=1}^{J,K}$ ($G=1$, $d_1=JK$). \label{simulations-huge}}
\begin{tabular}{lrrrrrrr}
 & RMSE$\times 10^{-4}$ & ESE$\times 10^{-4}$ & ASE$\times 10^{-4}$ & B$\times 10^{-6}$ & CI & L$\times 10^{-4}$ & ERR \\ 
Intercept & 2.26 & 2.25 & 1.99 & $-21.03$ & 0.93 & 7.81 & 0.07 \\ 
  $\bX_1$ & 0.35 & 0.35 & 0.31 & $-0.15$ & 0.92 & 1.22 & 0.08 \\ 
  $\bX_2$ & 0.35 & 0.35 & 0.31 & 1.07 & 0.93 & 1.22 & 0.07 \\ 
\end{tabular}
\end{table}
\newline In the Supplementary Material \citep{Hector-Song-DIQIF}, a quantile-quantile plot of the chi-squared statistic from Theorem \ref{thm:test} in the second set of simulations illustrates its appropriate asymptotic distribution. A quantile-quantile plot of the $Q$ statistic in \eqref{Q-statistic} in the linear regression setting comparing two partitions, one integrating over cohort only and one integrating over all data sources, is also given in the Supplementary Material \citep{Hector-Song-DIQIF}.

\section{METSIM analysis}
\label{sec:data}

\begin{figure}[h!]
\begin{center}
\parbox{\textwidth}{\caption{Estimated smoking effect for the xenobiotics pathway from the heterogeneous and integrative models categorized by significance at the $0.05/8$ level. \label{comparison-8}}}
\includegraphics[width=\textwidth]{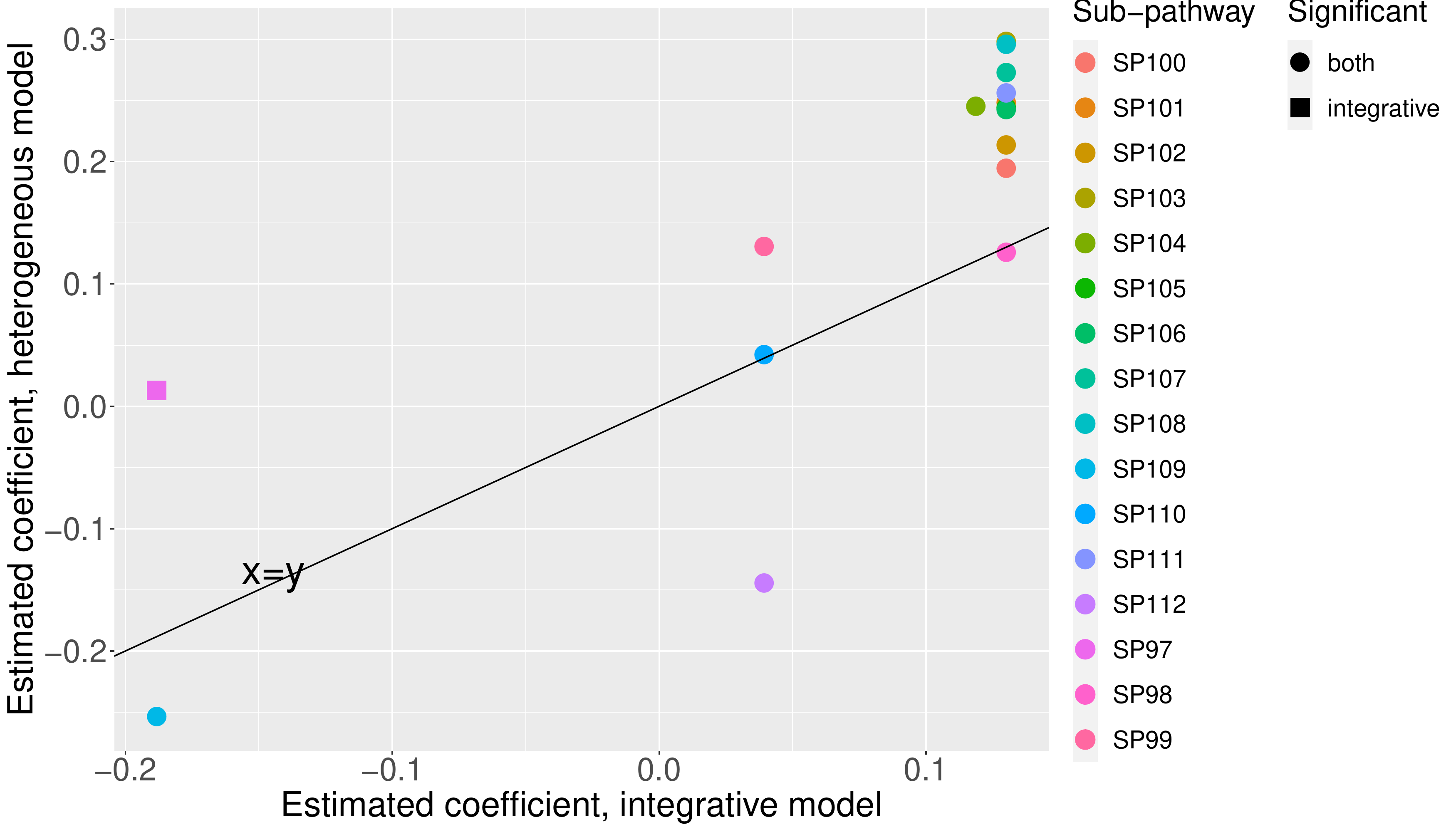}
\end{center}
\end{figure}

\begin{figure}[H]
\parbox{\textwidth}{\caption{Estimated regression parameters for the xenobiotics pathway from the heterogeneous and integrative models. \label{parameters-8}}}
\begin{center}
\begin{subfigure}{\linewidth}
\centering
\caption{Heterogeneous model results.}
\includegraphics[width=\linewidth]{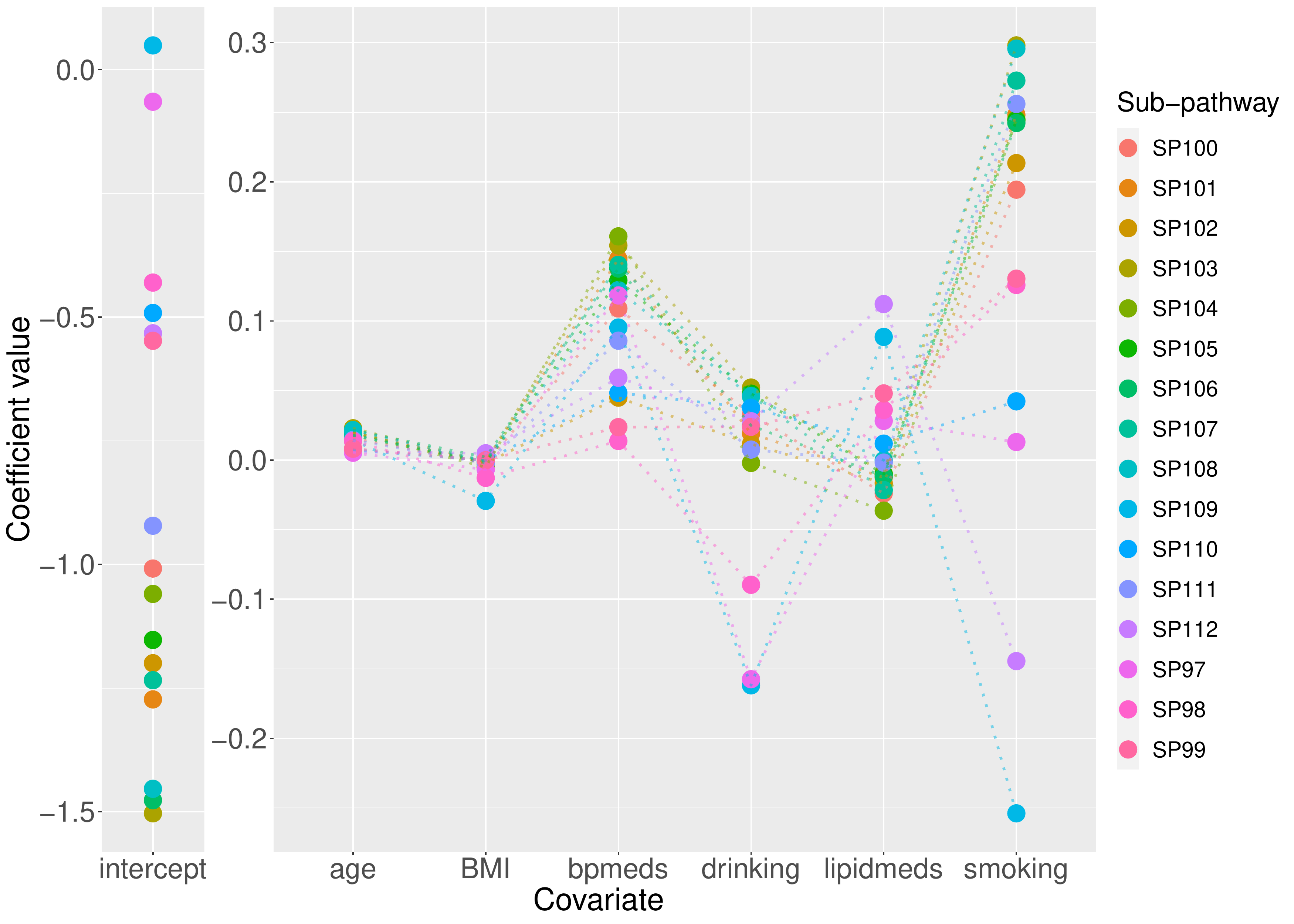}
\end{subfigure}%
\newline
\begin{subfigure}{\linewidth}
\caption{Integrative model results.}
\includegraphics[width=\linewidth]{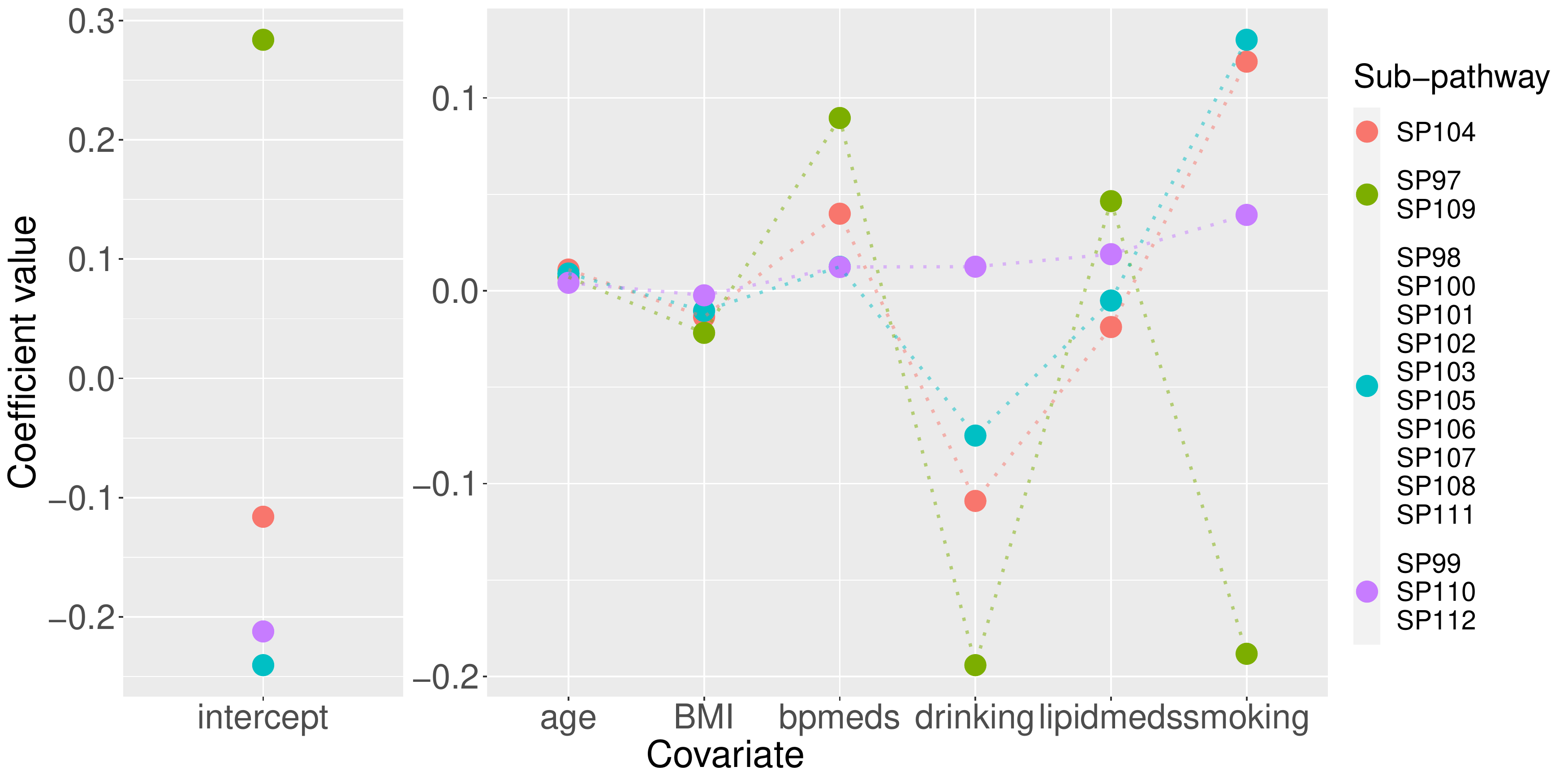}
\end{subfigure}
\end{center}
\end{figure}

We illustrate the application of the proposed method to the integrative analysis of metabolic pathways in the METSIM study described in Section \ref{sec:METSIM}. The METSIM study profiled $N=6223$ men in $K=4$ separate samples with sample sizes $n_1=1229$, $n_2=2950$, $n_3=1045$ and $n_4=999$. They measured $1018$ metabolites belonging to $112$ sub-pathways grouped in eight pathways with distinct biological functions. For each pathway, we investigate the association between smoking and metabolites in the pathway using our distributed and integrated quadratic inference functions approach to account for heterogeneity and correlation in metabolite sub-pathways. 
\\
Consider pathway $s \in \{1, \ldots, 8\}$. To illustrate the statistical efficiency gains from accounting for correlation between sub-pathways and combining models over independent cohorts, we first estimate sub-pathway and cohort specific effects and integrate them over cohorts (but not over sub-pathways); we then selectively combine regression models across sub-pathways based on a known partition $\mathcal{P}$. More specifically, we first estimate a heterogeneous model with partition $\mathcal{P}^{h}=\{ \mathcal{P}^{h}_j \}_{j=1}^{J}$, $\mathcal{P}^{h}_j=\{(j,1), \ldots, (j,K) \}$ that yields unique values of the regression coefficients for each sub-pathway. We then create a partition $\mathcal{P}^{i}$ of $\mathcal{P}^{h}$ with cardinality $G$ by selectively combining sub-pathways based on prior knowledge and estimate an integrative model. Details on the combination scheme can be found in the Supplementary Material \citep{Hector-Song-DIQIF} along with plots of parameters estimates. Note that the energy pathway is only constituted of two sub-pathways which cannot be combined. \\
We describe the marginal model for metabolites in pathway $s$. Denote by $J$ the number of sub-pathways and $M$ the number of metabolites in pathway $s$. Let $y_{ir,jk}$ denote the value of metabolite $r \in \{ 1, \ldots, m_j \}$ in sub-pathway $j\in \{1, \ldots, J \}$ for participant $i \in \{1, \ldots, n_k \}$ in cohort $k \in \{1, \ldots, K \}$, and let $\by_{jk}=(y_{ir,jk})_{i,r=1}^{n_k,m_j}$. Consider the marginal regression model
\begin{equation}
\begin{aligned}
E(y_{ir,jk})=&\theta_{jk,0} + \theta_{jk,1}smoking_{i,k} + \theta_{jk,2} age_{i,k} + \theta_{jk,3} BMI_{i,k} + \theta_{jk,4} drinking_{i,k} + \\
& \theta_{jk,5} bpmeds_{i,k} + \theta_{jk,6}lipidmeds_{i,k}, \label{data-model}\\
i=1, \ldots, n_k,& ~r=1, \ldots, m_j, ~j=1, \ldots, J, ~k=1, \ldots, 4, 
\end{aligned}
\end{equation}
where $smoking_i$ is participant $i$'s smoking status ($0$ for non-smoker, $1$ for smoker), $age_{i.k}$ is participant $i$'s age (range: $45.3$ to $74.4$ years), $BMI_{i,k}$ is participant $i$'s BMI (range: $16.9$ to $55.4$ kg/m\textsuperscript{2}), $drinking_{i,k}$ is an indicator for participant's $i$'s alcohol consumption ($0$ for non-consumer, $1$ for consumer), $bpmeds_{i,k}$ is an indicator for participant $i$'s blood pressure medication use ($0$ for no use, $1$ for use), and $lipidmeds_{i,k}$ is an indicator for participant $i$'s lipid medication use ($0$ for no use, $1$ for use), at the time of data collection. Let $\btheta_{jk}=(\theta_{0,jk}, \dots, \theta_{6,jk})$. Plots and tables of regression parameter estimates from heterogeneous and integrative models for all pathways, and comparisons of significance levels of the smoking effect between the two partitions, are given in the Supplementary Material \citep{Hector-Song-DIQIF}. In our discussion of the xenobiotics pathway, we use the following pathway short names: food component/plant (SP97); drug - other (SP98); xanthine metabolism (SP99); chemical (SP100); drug - analgesics, anesthetics (SP101); benzoate metabolism (SP102); tobacco Metabolite (SP103); drug - topical agents (SP104); drug - antibiotic (SP105); drug - cardiovascular (SP106); drug - neurological (SP107); drug - respiratory (SP108); drug - psychoactive (SP109); drug - gastrointestinal (SP110); bacterial/fungal (SP111); drug - metabolic (SP112).\\
The statistics $Q^{h}_N(\widehat{\btheta}^{h})$ and $Q^{i}_N(\widehat{\btheta}^{i})$ in Theorem \ref{thm:test} for heterogeneous and integrative models respectively and the GMM-BIC values from \eqref{GMM-BIC} are reported in Table \ref{table-Q}. 
\begin{table}[ht]
\centering
\caption{Statistics and GMM-BIC for heterogeneous and integrative models, by pathway. \label{table-Q}}
\resizebox{\textwidth}{!}{
\begin{tabular}{lrrrrrrr}
pathway & $M$ & $J$ & $G$ & $Q^{h}_N(\widehat{\btheta}^{h})$ (d.f.) & $Q^{i}_N(\widehat{\btheta}^{i})$ (d.f.) & $BIC(\mathcal{P}^{h})$ & $BIC(\mathcal{P}^{i})$ \\
amino acid & 203 & 15 & 8 & 803.69 (315) & 1070.98 (364) & $-1948.15$ & $-2108.93$ \\ 
  carbohydrate & 24 & 5 & 4 & 487.44 (105) & 531.04 (112) & $-429.84$ & $-447.39$ \\ 
  cofactors and & \multirow{2}{*}{34} & \multirow{2}{*}{7} & \multirow{2}{*}{4} & \multirow{2}{*}{354.32 (147)} & \multirow{2}{*}{526.06 (168)} & \multirow{2}{*}{$-929.87$} & \multirow{2}{*}{$-941.59$} \\ 
  \multicolumn{1}{r}{vitamins} & & & & & & & \\
  energy & 9 & 2 & 2 & 162.01 (42) & 162.01 (42) & $-204.9$ & $-204.9$ \\ 
  lipid & 465 & 54 & 3 & 2050.73 (1134) & 4051.8 (1491) & $-7855.9$ & $-8973.59$ \\ 
  nucleotide & 35 & 7 & 3 & 453.26 (147) & 644.2 (175) & $-830.93$ & $-884.6$ \\ 
  peptide & 36 & 6 & 4 & 858.05 (126) & 938.33 (140) & $-242.69$ & $-284.71$ \\ 
  212 & 16 & 4 & 1325.77 (336) & 1890.46 (420) & $-1609.53$ & $-1778.66$ \\   
\end{tabular}
}
\end{table}
Values of the BIC criterion support the choice of integrative partitions for each pathway. We also highlight the reduction in parameter dimension obtained from selectively combining models. For example, the heterogeneous model for the xenobiotics pathway has $16\times 7=112$ regression parameters, and the integrative model has $4\times 7=28$. This is due to the combination of the following sub-pathways: SP97 and SP109; SP98, SP100, SP101, SP102, SP103, SP105, SP106, SP107, SP108 and SP111; SP99, SP110 and SP112. Only SP104 has a sub-pathway specific regression model. This reduction in dimension results in the greater significance of the smoking effect in the food component/plant (SP97) sub-pathway in the integrative partition compared to the heterogeneous partition, as seen in the comparison of significance levels of the smoking effect between the two partitions in Figure \ref{comparison-8}.
\newline Regression parameter estimates for the xenobiotics pathway are displayed in Figure \ref{parameters-8}. Based on the integrative models, we find that the effect of smoking is significant in multiple sub-pathways of the xenobiotics pathway. The tobacco metabolite sub-pathway (SP103) is combined with multiple sub-pathways. The estimated effect of smoking in this integrated sub-pathway is $0.13$ with a standard error of $0.011$ and $p$-value $1.2\times 10^{-30}$. We observe from Figure \ref{comparison-8} that by combining sub-pathways in the integrative model we are able to borrow information across sub-pathways and obtain more precise inference.

\section{Discussion}
\label{sec:discussion}

The proposed method can be viewed as a generalization of both seemingly unrelated regression and meta-analysis, striking a balance between the two that leverages correlation and partial homogeneity of regression equations. The distributed quadratic inference approach is privacy-preserving and computationally appealing because data source analyses can be run simultaneously in parallel and only one round of communication is necessary to compute the integrated estimator in \eqref{def:one-step-estimator-both}. The test of model fit proposed in Theorem \ref{thm:test} and the $\chi^2$ test of homogeneity in \eqref{Q-statistic} are derived from the unique properties of the generalised method of moments and provide a principled approach to model building that is lacking with other state of the art correlated data analysis techniques, such as generalised estimating equations. \\
The selective combination scheme over a data source partition has also been studied in \cite{Wang-Wang-Song-2016} and \cite{Tang-Song}. While we require specification of the data partition, their meth-
\noindent ods learn the partition in a data-driven way, which can be advantageous. Inference with the fused lasso, however, is burdened by debiasing methods that can be ill-defined or computationally burdensome. Additionally, the fused lasso approaches do not provide a formal procedure to test the validity of the parameter fusion scheme, relying instead on visualization such as dendograms. Our method is clearly advantageous when an approximately known partition exists.\\
Limitations of the proposed method include the need for a pre-defined partition of data sources defining regions of parameter homogeneity, which typically is given by related scientific knowledge but may occasionally be lacking in practice. Data pre-processing and learning and the test in \eqref{Q-statistic} may help in determining an appropriate partition. Additionally, standard errors tend to be underestimated in small sample sizes or when the dimension of the moment conditions is large; this phenomenon has been well documented in the generalised method of moments (see \cite{Hansen-Heaton-Yaron} and others in the same issue).

\section*{Acknowledgements}

We are very grateful to Drs. Michael Boehnke and Markku Laakso for their generosity in letting us use their data.  We are grateful to Dr. Lan Luo for sharing her code to estimate parameters using quadratic inference functions. This work was supported by grants R01ES024732 and NSF1811734.

\appendix

\section{Assumptions}
\label{appendix:conditions}

\begin{enumerate}[label=(A.\arabic*), labelindent=0pt]
\item \label{QIF-conds}
Assumptions for consistency and asymptotic normality of $\widehat{\btheta}_{jk}$ for data source $(j,k)$ in $g$th partition set $\mathcal{P}_g$:
\begin{enumerate}[label=(\roman*), labelindent=0pt]
\item $\bC^{-1}_{jk}$ is positive definite and $\bC^{-1}_{jk}E_{\btheta_{0,g}}\{ \bpsi_{i,jk}(\btheta_{jk})\}=\boldsymbol{0}$ if and only if $\btheta_{jk}=\btheta_{0,g}$
\item the true value $\btheta_0=(\btheta_{0,g})_{g=1}^G$ is an interior point of $\Theta=\times_{g=1}^G \Theta_g$, and $\Theta_g$ are compact
\item $\bpsi_{i,jk}(\btheta_{jk})$ is continuous at each $\btheta_{jk}$ with probability one
\item $E_{\btheta_0} \{ \sup_{\btheta_{jk} \in \Theta_g} \left\| \bpsi_{i,jk}(\btheta_{jk}) \right\| \}<\infty$
\item $\bpsi_{i,jk}(\btheta_{jk})$ is continuously differentiable in a neighborhood $\mathcal{N}$ of $\btheta_{0,g}$ with probability approaching one
\item $E_{\btheta_{0,g}}\{ \bpsi_{i,jk}(\btheta_{0,g})\} =\boldsymbol{0}$ and $E_{\btheta_{0,g}} \{ \left\| \bpsi_{i,jk}(\btheta_{0,g}) \right\|^2\}$ is finite and positive-definite
\item $E_{\btheta_0} \{ \sup_{\btheta_{jk} \in \mathcal{N}} \left\| \nabla_{\btheta_{jk}} \bpsi_{i,jk}(\btheta_{jk}) \right\| \}<\infty$;  $[E_{\btheta_0} \{ \nabla_{\btheta_{0,g}} \bpsi_{i,jk}(\btheta_{0,g})\}]^T \bC^{-1}_{jk} E_{\btheta_0}\{ \nabla_{\btheta_{0,g}} \allowbreak \bpsi_{i,jk}(\btheta_{0,g}) \}$ is nonsingular
\end{enumerate}
\item \label{DDIMM-conds} 
For any $\delta_N \rightarrow 0$, 
\begin{align*}
\sup \limits_{\left\| \btheta - \btheta_0 \right\| \leq \delta_N} \frac{N^{1/2}}{1+N^{1/2} \left\| \btheta-\btheta_0 \right\| } \left\| \bPsi_N(\btheta) - \bPsi_N(\btheta_0) - E_{\btheta_0} \bPsi_N(\btheta) \right\| =O_p(N^{-1/2}).
\end{align*}
\end{enumerate}

\section{Proof of Theorem \ref{thm:efficiency}}
\label{appendix:proofs}

Let $\bv_{jk}(\btheta_{jk})=Var_{\btheta_{0,g}}(\bpsi_{i,jk}(\btheta_{jk}))$ . From assumption \ref{QIF-conds} and Theorem \ref{thm:norm}, $Avar(\sqrt{n_k} \widehat{\btheta}_{jk}) = \{ \bs_{jk}(\btheta_{0,g}) \bv^{-1}_{jk}(\btheta_{0,g}) \bs_{jk}^T(\btheta_{0,g}) \}^{-1}$, and
\begin{gather*}
Avar(\sqrt{N} \widehat{\btheta}) =\left\{ \bs^T(\btheta_0) \bv^{-1}(\btheta_0) \bs(\btheta_0) \right\}^{-1}, \quad Avar(\sqrt{N} \widehat{\btheta}^g)=\left[ \left\{ \bs^T(\btheta_0) \bv^{-1}(\btheta_0) \bs(\btheta_0) \right\}^{-1} \right]_g,
\end{gather*}
where $\left[ \bA \right]_g$ denotes the submatrix of a matrix $\bA$ consisting of rows and columns corresponding to data sources in $\mathcal{P}_g$. Let $[ \bv(\btheta) ]_{(j,k)}$ denote submatrix of $\bv(\btheta)$ consisting of rows and columns corresponding to data source $(j,k)$, and let $[\bv(\btheta)]_{-(j,k);}$, $[\bv(\btheta)]_{;-(j,k)}$ and $[\bv(\btheta)]_{-(j,k)}$ denote the submatrices of $\bv(\btheta)$ eliminating respectively rows, columns, and rows and columns corresponding to data source $(j,k)$. Clearly $(n_k/N)\bv_{jk}(\btheta_{jk})$ is a submatrix of $\bv(\btheta)$: $(n_k/N) \bv_{jk}(\btheta_{jk})=\left[ \bv(\btheta) \right]_{(j,k)}$.\\
Consider $(j,k)=(1,1)$, which is in partition set $\mathcal{P}^{g_1}$ for some $g_1 \in \{1, \ldots, G\}$ (this is without loss of generality since we can reorganize the rows of $\bpsi_i$ for $(j,k) \neq (1,1)$ to make $(j,k)=(1,1)$). We write
\begin{align*}
\bv(\btheta)&=\left( \begin{array}{ll}
[\bv(\btheta)]_{(1,1)} & [\bv(\btheta)]_{;-(1,1)} \\ \relax
[\bv(\btheta)]_{-(1,1);} & [\bv(\btheta)]_{-(1,1)}
\end{array} \right).
\end{align*}
By Corollary 7.7.4. in \cite{Horn-Johnson}, 
\begin{align*}
\bs_{11}(\btheta_{g_0,1}) \bv^{-1}_{11}(\btheta_{g_0,1}) \bs^T_{11}(\btheta_{g_0,1}) &\prec \bs_{11}(\btheta_{g_0,1}) \frac{n_1}{N} \left[ \bv^{-1}(\btheta_0) \right]_{(1,1)}  \bs_{11}^T(\btheta_{g_0,1})
\end{align*}
where $\preceq$ denotes L\"{o}wner's partial ordering in the space of nonnegative definite matrices. By the definition of $\bs_{g_1}(\btheta_{g_0,1})$ in Section \ref{subsec:theory}, this implies that
\begin{align*}
\left\{ \bs^T_{g_1}(\btheta_{g_0,1}) \left[ \bv^{-1}(\btheta_0) \right]_{g_1} \bs_{g_1}(\btheta_{g_0,1}) \right\}^{-1} &\prec \left\{ \frac{n_1}{N} \bs_{11}(\btheta_{g_0,1}) \bv^{-1}_{11}(\btheta_{g_0,1}) \bs^T_{11}(\btheta_{g_0,1})\right\}^{-1}\\
&=\lim \limits_{n_1 \rightarrow \infty} \frac{N}{n_1} Avar(\sqrt{n_1}\widehat{\btheta}_{11}).
\end{align*}
Again by Corollary 7.7.4. in \cite{Horn-Johnson}, we have that
\begin{align*}
Avar(\sqrt{N} \widehat{\btheta}^{g_1})=\left[ \left\{ \bs^T(\btheta_0) \bv^{-1}(\btheta_0) \bs(\btheta_0) \right\}^{-1} \right]_{g_1} & \prec \left\{ \bs^T_{g_1}(\btheta_{g_0,1}) \left[ \bv^{-1}(\btheta_0) \right]_{g_1} \bs_{g_1}(\btheta_{g_0,1}) \right\}^{-1} ,
\end{align*}
implying $Avar(\sqrt{N} \widehat{\btheta}^{g_1}) \preceq \{ \lim_{n_k \rightarrow \infty} (N/n_k) \} Avar(\sqrt{n_1} \widehat{\btheta}_{11})$.
\qed

\bibliographystyle{apalike} 
\bibliography{DIQIF-bib-20190125}

\end{document}